\documentclass[conference]{IEEEtran}
\IEEEoverridecommandlockouts
\usepackage{cite}
\usepackage{amsmath,amssymb,amsfonts}
\usepackage{algorithmic}
\usepackage{graphicx}
\usepackage{textcomp}
\usepackage{xcolor}
\usepackage{comment}
\usepackage{enumitem}
\usepackage{balance}
\usepackage{graphicx}
\usepackage{grffile}
\usepackage{subcaption}
\usepackage{float}
\usepackage{amssymb}
\usepackage{balance}

\def\BibTeX{{\rm B\kern-.05em{\sc i\kern-.025em b}\kern-.08em
    T\kern-.1667em\lower.7ex\hbox{E}\kern-.125emX}}
\begin{document}

\title{%Feature-Sniffer: An Access-Point Tool for Real-Time Network Traffic Features Capturing
Feature-Sniffer: Enabling IoT Forensics in OpenWrt based Wi-Fi Access Points
}
\author{

    \IEEEauthorblockN{Fabio Palmese}
    \IEEEauthorblockA{
        \textit{DEIB, Politecnico di Milano} \\
        Milan, Italy\\
        fabio.palmese@polimi.it}
    \and

    \IEEEauthorblockN{Alessandro E. C. Redondi}
    \IEEEauthorblockA{
        \textit{DEIB, Politecnico di Milano} \\
        Milan, Italy\\
        alessandroenrico.redondi@polimi.it
        }
    \and

    \IEEEauthorblockN{Matteo Cesana}
    \IEEEauthorblockA{
        \textit{DEIB, Politecnico di Milano} \\
        Milan, Italy\\
        matteo.cesana@polimi.it
    }

}

\begin{comment}
\author{
\IEEEauthorblockN{Author name}
\IEEEauthorblockA{\textit{Author Affiliation} \\
Affiliation city\\
Author Contacts
}
}
\end{comment}
\maketitle

\begin{abstract}
%Lorem ipsum dolor sit amet, consectetur adipiscing elit. Quisque nec condimentum diam. Etiam a pretium arcu. In pharetra, odio eget congue convallis, nulla erat malesuada est, a porttitor enim nibh ut risus. Fusce leo lorem, volutpat vitae lacinia sodales, condimentum ornare sem. Mauris leo velit, pharetra ac magna eget, consequat efficitur tellus. Fusce rhoncus nec dolor et feugiat. Nulla bibendum mauris et enim semper pellentesque.
%Suspendisse potenti. Nunc in mi est. Curabitur tincidunt accumsan suscipit. Nam posuere pellentesque rhoncus. Nullam sed elementum quam. Nullam sed ex turpis. Praesent vel sollicitudin ligula. Morbi a finibus diam. Nunc at ipsum iaculis nisi fermentum vestibulum eu sit amet leo. Donec mattis nec felis et ornare. Sed vitae facilisis lacus, in facilisis libero. Ut sed mauris lorem. Phasellus efficitur leo a velit euismod sodales. Etiam mattis nunc tellus, ac faucibus tortor bibendum a. Lorem ipsum dolor sit amet, consectetur adipiscing elit. Quisque nec condimentum diam. Etiam a pretium arcu. In pharetra, odio eget congue convallis, nulla erat malesuada est.
The Internet of Things is in constant growth, with millions of devices used every day in our homes and workplaces to ease our lives. Such a strict coexistence between humans and smart devices makes the latter digital witnesses of our every-day lives through their sensor systems. This opens up to a new area of digital investigation named IoT Forensics, where digital traces produced by smart devices (network traffic, in primis) are leveraged as evidences for forensic purposes. It is therefore important to create tools able to capture, store and possibly analyse easily such digital traces to ease the job of forensic investigators. This work presents one of such tools, named Feature-Sniffer, which is thought explicitly for Wi-Fi enabled smart devices used in Smart Building/Smart Home scenarios. Feature-Sniffer is an add-on for OpenWrt-based access points and allows to easily perform online traffic feature extraction, avoiding to store large PCAP files. We present Feature-Sniffer with an accurate description of the implementation details, and we show its possible uses with practical examples for device identification and activity classification from encrypted traffic produced by IoT cameras. We release Feature-Sniffer publicly for reproducible research. %The goal of the work has been achieved, showing the feasibility of the two application cases with an effort-less dataset construction using the tool we propose.
%For these reasons it is becoming important to keep such devices traces for forensics purposes, traces not easy to be preserved given the devices constraints. Considering that the main traces of such devices relay on the produced network traffic, this works presents Feature-Sniffer an open-source tool enabling IoT Forensics in WiFi access points. The tool allows to easily perform online traffic feature extraction for devices connected to the access point, avoiding to store large PCAP files. After a presentation of the tool with an accurate description of the implementation details, this work shows its possible uses with practical examples for device identification and activity classification. The goal of the work has been achieved, showing the feasibility of the two application cases with an effort-less dataset construction using the tool we propose.

\end{abstract}


\section{Introduction}
\label{sec:intro}
%Bla bla on IoT devices in smart homes / WLANS
The Internet of Things is moving from a visionary concept to reality, revolutionizing our lives. The number of smart, connected devices surrounding us is increasing everyday, thanks to innovations in low-power communication technologies, sensor systems, energy-efficient microcontrollers and electric battery devices. Among the several application scenarios of the IoT, Smart Buildings and Smart Homes constitute one of the most investigated. Being rich of both energy resources (the power grid) and connectivity features (WiFi, most of the times), buildings and homes are the perfect playground for IoT prototypes of any kind, also considering that they constitute the places where humans spend most of their time.
%Not only digital forensics, but also actual forensics
Such a strict coexistence between humans and smart devices creates a twofold effect: on the one hand, it opens the way to a multitude of novel cyber-attacks, focused e.g., on disrupting security systems of a home/building or stealing personal information from IoT devices. On the other hand, it makes such devices witnesses of our every-day lives through their sensor systems. For these reasons, in the last few years, the term ``IoT Forensics'' has been coined in order to refer to all those activities related to gathering sensitive information from IoT devices, whether they have been compromised or not \cite{iot_forensics_survey}.

%Need for tools to analyse and store traffic, possibly before it is hidden via NAT/gateways
IoT forensics is different and more challenging from traditional digital forensics. First, IoT devices have generally limited or no permanent memory to analyse during an investigation. Therefore, the main asset that could be leveraged for a forensic analysis is the network traffic they produce and exchange with other entities (e.g., cloud services). 
However, network traffic is of ephemeral and transient nature: as a mere mean of communication, it
ceases to exist after it is received by the intended recipient. Therefore, to enable a-posteriori forensic
analysis based on network traffic, it is mandatory to capture and store it in an efficient manner. It is hence imperative to put in place systems able to easily acquire, store and analyse such network traffic as close as possible to where it is produced, in order to avoid loss of information.
As an example, IoT traffic from multiple devices connected to an access point using NAT technology (as it is often the case) may be of difficult analysis when observed from the external, since all devices appear with the same source MAC and IP addresses. This makes it difficult to perform per-device, feature-based traffic analysis, which is the de-facto approach in case of encrypted traffic such as the one produced by current IoT devices~\cite{iot_dataset}.
Moreover, the growing presence of local IoT traffic (e.g., control traffic from a home assistant or smartphone application to an IoT device in the same LAN) calls for  capturing devices which are themselves part of the local networks under study.
At the same time, storing the complete network packets produced by IoT devices (e.g., in form of PCAP files) may be infeasible on the long run, due to the large storage space required.
For all these reasons this work proposes Feature-Sniffer, an open-source tool enabling IoT forensics on any WiFi access point supporting the OpenWrt project. 
Feature-Sniffer allows to easily perform online per-device traffic feature extraction and storage through the access point graphical web interface, making the network forensic-ready in a click. The tool allows to set-up easily IoT forensics analysis activities spanning long periods of time with no effort, avoiding the unnecessary storage of large PCAP files.
The contribution of this paper is as follows: first we describe Feature-Sniffer, focusing on its working principle as well as on the implementation details. After an accurate description of the tool and an overview on the performance impact on the network, we use Feature-Sniffer to perform two IoT forensics activities related to IoT video cameras: (i) device identification and (ii) activity recognition from encrypted traffic. Finally, we release Feature-Sniffer publicly\footnote{https://github.com/fpalmese/feature-sniffer}.\\
This paper's remainder is organized as follows: Section \ref{sec:relwork} presents the main related works on IoT Forensics. Section \ref{sec:tool} provides an accurate description of the Feature-Sniffer tool with implementation, installation and performance details. Two possible application cases are discussed in Section \ref{sec:application} whose experimental results obtained with the application of Machine Learning algorithms are summarized in Section \ref{sec:results}. Section \ref{sec:concl} concludes the works with final remarks and highlights future research directions.

\section{Related Work}
\label{sec:relwork}

In the last years, with the increase of interest in IoT and in Digital Forensics, several works focused on the study of IoT devices behaviour as possible evidence for forensic investigations. In particular the literature on IoT Forensics can be roughly divided into three categories: IoT devices identification/classification, real-life event detection through IoT device activity recognition and general IoT Forensics frameworks. In the first category, a large number of works has widely explored the use of network traffic traces from IoT devices for device classification. An accurate overview on the main related works is reported in \cite{survey_iot_classification} where the authors present a complete survey on the topic of IoT device fingerprinting using several source of traces ranging from network traffic to hardware characteristics. In the field of device identification relying on network traffic, the work in \cite{iot_dataset} contributes to the research by proposing an accurate method to fingerprint 31 IoT and non-IoT devices applying machine learning techniques to network traffic features achieving promising results in terms of accuracy. In addition, the authors contribute by publicly releasing the dataset used in their experiments (made of 1 PCAP/CSV file per day for 60 days) to be accessible for research purposes.
Regarding real-life activity identification, several works have analyzed different IoT devices and applied techniques for different use cases (e.g. \cite{iot_athena, information_exposure, echo_classification}).
%Regarding real-life activity identification, several works have analyzed different IoT devices and applied techniques for different use cases. The authors of \cite{iot_athena} developed a system based on two algorithms able to unveil sequences of IoT devices activities from raw network traffic contained in PCAP files. Their system called IoTAthena can be used to characterize signatures of IoT device activities and hence use them for forensics purposes i.e., detecting anomalies/cyber attacks or classifying user behaviour concatenating activities detected from different devices.
%The work in \cite{information_exposure} quantifies the information exposure from IoT devices across different networks, geographic regions and user interactions applied to 81 different devices located in two environments in the US and UK. The authors show that even if the communication relies on encrypted traffic, private information can be gathered from network traffic produced from such devices and hence the devices activities can be unveiled.
%In the particular case of Home Assistant devices, the work in \cite{echo_classification} presents how to use machine learning approaches applied to network traffic produced from Amazon Echo devices to classify the questions being performed from the user to the Home Assistant, distinguishing with over 97\% accuracy six different question categories.
We report here the main works regarding the particular case of video camera network traffic classification for user activity recognition. 
%To conclude the related works on real-life user behaviour identification, we report the main works regarding the particular case of video camera network traffic classification for user activity recognition.
The authors of \cite{video_classification} have analyzed the traffic produced by two different IP cameras to classify 9 different user daily activities performed in front of the camera (dressing, sporting, reading etc...). The results have been obtained by applying six machine learning classifier applied to several statistical features extracted from the network traffic produced by the cameras, stored in 9000 PCAP files each containing only one user activity. The main considered features are packet inter-arrival times, transmission rates and packet lengths, which are then fed into a Random Forest classifier model.
The work in \cite{wampler2015information} shows that it is possible to detect motion or scene change from encrypted traffic produced by IP cameras regardless of the codec used by the devices, performing a graphical analysis plotting over time several traffic features including arrival time between packets, packet sizes, and video stream bandwidth.
In \cite{li2016side} the authors use two common IoT cameras (Google and Samsung) to demonstrate how attackers can easily infer basic activities of daily living (4 activities are distinguished) based only on the traffic size statistics of an encrypted video stream.

%Finally, very few works have presented practical IoT Forensics frameworks like the one presented in this paper. It is worth to report a work from the far 2010 \cite{net_forensics_acquisition}, in which the authors, propose a tool to automatically collect network traffic connection logs from OpenWrt based Access Points using \texttt{tcpdump} and to send the results to the network administrator periodically via email.

Finally, very few works have presented practical IoT Forensics frameworks like the one presented in this paper. It is worth to report a work from 2010 \cite{net_forensics_acquisition}, in which the authors propose a tool that allows to automatically collect connection logs in OpenWrt based Access Points and transmit them periodically via e-mail.

\section{Feature-Sniffer}
\label{sec:tool}

\begin{figure}[t]
    \begin{center}
      \includegraphics[width=8cm]{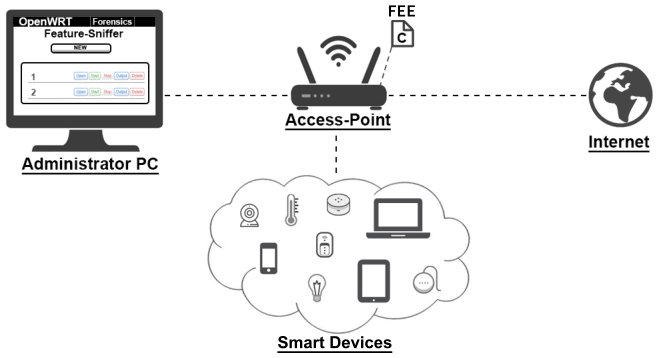}
      \caption{Feature-Sniffer architecture. The OpenWrt control panel on the access point allows to interact with the feature extraction engine in order to perform IoT Forensics analysis}
      \label{fig:architecture}
    \end{center}
\end{figure}

\subsection{Overview}
This paper proposes Feature-Sniffer, a tool which allows to compute network traffic features in an online fashion on any WiFi access point operated by the OpenWrt firmware, hence avoiding the cumbersome tasks of (i) dumping network traffic to PCAP files and (ii) implementing ad-hoc routines for analysing the captured traffic.  %to be divided in the two parts of capturing and parsing/analysis.
The tool is designed to be extremely user friendly, so that network administrators, data scientists or forensic investigators can access to a multitude of network traffic features with easy steps, instead of building ad-hoc traffic capture and analysis pipelines. 
%By using the tool we propose, a network administrator, data scientist or forensics investigator can extract features from a PCAP file or from Live network traffic with fast and easy steps, instead of building its own program to parse the PCAP and calculate the features. This eases the computation and helps to save time (write code for parsing + computation) and space (storage).
In a nutshell, Feature-Sniffer analyzes all network packets that flow through the Access Point and organizes them in time windows of user-defined duration. For each time window and for each device connected to the Access Point (identified by its MAC address), the tool computes and saves a set of traffic features selected by the user through a web Graphical User Interface (GUI).
Figure \ref{fig:architecture} sketches the architecture of the proposed tool, which is composed of two main parts: (i) an application for the LuCI web interface, developed to access all functionalities directly from the access point control panel; (ii) a feature extraction engine, developed in the C programming language and easily configurable through the web interface. The following sections give details on these two main components. %For this purpose a new "Forensics" section is introduced in the Web GUI to let the user edit easily the capture settings and manage the configurations.
%For the traffic capture we have developed a C program (described later) to be installed in the Access Point and that is started with the parameters selected by the user.
%Feature-Sniffer is publicly available for research purposes\footnote{https://github.com/fpalmese/feature-sniffer}.

\begin{figure*}[t]
    \begin{center}
      \includegraphics[width=1.28\columnwidth]{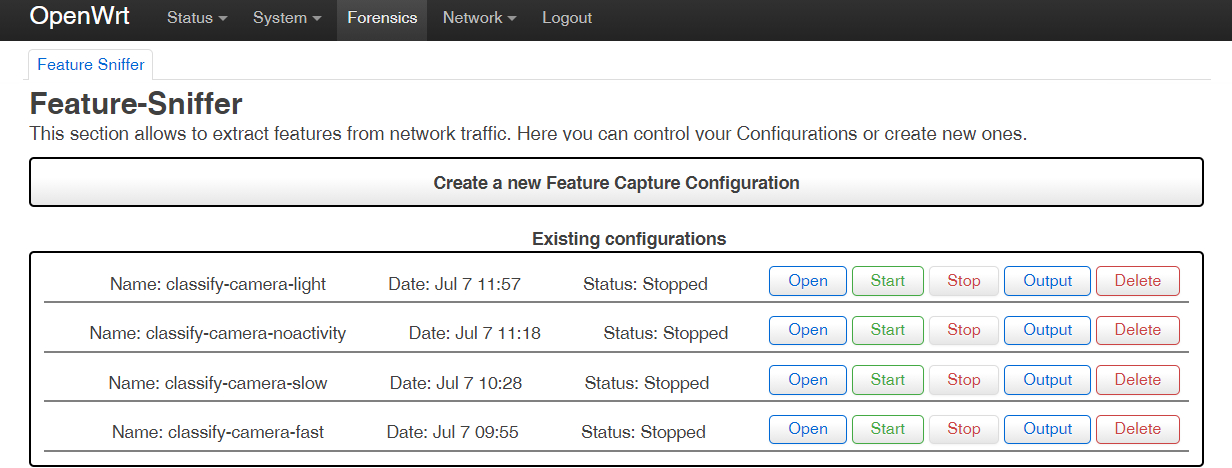}
      \caption{Feature-Sniffer GUI homepage. The user can control existing configurations or create new configurations.}
      \label{fig:feature-sniffer-home}
    \end{center}
\end{figure*} 
  
\subsection{Web Interface}
The Feature-Sniffer user interface is an add-on built on top of LuCI functionalities\footnote{https://github.com/openwrt/luci}. LuCI is the standard web user interface for OpenWrt systems since 2008. It is based on the Lua programming language, which easily allows the creation of customized extensions. 
As illustrated in Figure \ref{fig:feature-sniffer-home}, we extended LuCI with a new user interface for IoT Forensics. The interface allows the network administrator to quickly create traffic analysis \textit{configurations}, in order to directly compute specific features from the network traffic flowing through the access point.
Each configuration is characterized by the following parameters:
\begin{itemize}[nosep, wide]
    \item \textit{Name and Description}: representing the configuration unique identifier as well as a brief optional text describing the configuration purposes.
    \item \textit{Window duration}: the duration in time of each analysis window. Traffic features are computed independently in each window.
    \item \textit{Capture filter}: an optional string following the \textit{pcap\_filter} syntax to extract features only from the traffic of interest (e.g., only outgoing UDP traffic).
    \item \textit{Device filter}: the user may insert here the MAC addresses of the devices to consider in the analysis. When empty, all connected devices are considered.

    %\item Capture Type: Live or from PCAP. The Live capture requires the user to specify the interface from which the packets will be captured, while the PCAP capture requires the user to specify the full path of the PCAP file to be parsed (that needs to be previously uploaded in the Access Point).

    \item \textit{Features list}: Feature-Sniffer allows the user to select which traffic features to compute for each time window and each associated device. The most popular features used in network traffic analysis are available, computed from several statistics regarding the packet size (72 features), the payload length (63 features) and the inter-arrival times (45 features). The available statistics include count, sum, mean, median, mode, variance , standard deviation and Kurtosis. Specific options for TCP and UDP packets can be selected, or both the protocols can be considered combined. The same holds for Downlink (DL) and Uplink (UL) packets. 
    %The complete set of features available in Feature-Sniffer is reported in Table \ref{table:features}. 
    Additionally, the user can select 144 features derived from the Probability Mass Function (PMF) of packet lengths, discretized in intervals of 100 bytes (i.e., containing the percentage of packets with length in range [0,100] bytes, [100,200] and so on...), again with the possibility of focusing on Uplink/Downlink and TCP/UDP packets.
    Finally we have included 5 more features containing the number of Remote IP addresses contacted by each device in the time window, the number of local TCP/UDP used ports and the number of remote TCP/UDP ports contacted. In total, Feature-Sniffer allows to compute up to 329 features per device in each time window.
    
    %a list of features to calculate and print for each time window of each device. Features contain statistical values (Count (\#), Sum ($\Sigma$), Mean ($\mu$), Median ($\mu \textsubscript{1/2}$), Mode (M), Variance ($\sigma$), Standard Deviation ($\sigma^2$), Kurtosis (K) ) of Packet/Payload sizes and of packet inter-arrival times. A distinction for TCP and UDP packets can be selected to split the features, or both the protocols can be considered combined. The same holds for Downlink (DL) and Uplink (UL) packets. We derived in total 72 features from Packet size, 63 features from Payload sizes and 45 features from inter-arrival times. A summary of these features is reported in table \ref{table:features}.
    %The user can also select 144 additional features derived from "Probability density functions" of packet lengths (containing the number of packets with length in range [0,100] bytes, [100,200] and so on...), dividing again Uplink/Downlink and TCP/UDP packets (16 features for each packet type in the first 8 rows of table \regf{table:features}).
    %Finally we have included 5 more features containing the number of Remote IP addresses contacted from the device in the time window, the number of local TCP/UDP used ports and the number of remote TCP/UDP ports contacted.
    %Considering all the features available, the user can select totally 329 features.
    
        \item \textit{Output parameters}: Feature-Sniffer allows the creation of customizable output, such as the addition of a ground-truth label for each device (useful, e.g., for device identification tasks), the use of feature headers name in the output, the possibility to create one file per each different associated device or the use of relative or absolute timestamps for each time window.
        
        %\begin{itemize}
        %\item Add labels for devices (specified by user)
        %\item Print feature headers in the output
        %\item Create one output per captured device
        %\item Use relative times for output (first window represents time 0)
        %\end{itemize}
\end{itemize}
%LuCI is t web interface written in Lua, easy to be extended with additional tools/applications and since 2008 it is the standard web user interface for OpenWrt based systems. 
%All the requests performed from the Feature-Sniffer user interface are handled in the web server, implemented using the LuCI server APIs to execute the correct actions for the program to work.
%We have developed an easy to use user interface that allows the network administrator to create one or more capture configurations in order to directly collect the interested features from the network traffic passing from the Access Point. 
The main page of the user interface shows all the existing configurations with the corresponding information and control buttons (start, stop, edit, download output, delete) as well as a section to create new configurations. Each configuration is saved in a \textit{.cfg} file, which is made compliant to the \texttt{libconfig} configuration.

\subsection{Feature Extraction Engine}
 Upon clicking on the start button, an HTTP request is issued to the web server on the access point which in turn will start the Feature Extraction Engine (FEE) according to the selected configuration. In particular, the FEE works as it follows: first, all configuration parameters are loaded with the help of libconfig library\footnote{https://github.com/hyperrealm/libconfig} API. Then, two concurrent threads are started: one thread is in charge of capturing packets for each associated device, using the \texttt{libpcap} library~\cite{libpcap} and grouping them in consecutive time windows according to the user defined window length and packet/device filters. When a window is ready, the thread pushes it into the input queue of a second thread, which is in charge of computing the selected features on the packets contained in the window and saving the result to a Comma Separated Value (CSV) file. We find this approach to be optimal in order to avoid losing packets due to the burden of feature computation.
 Whenever the user wants to interrupt the packet capture, it can use the stop control button to trigger a kill for the process. In particular to avoid losing some of the windows in the output, we have redefined the behaviour of the \texttt{SIGINT} handler: the signal triggers the break of the capture loop of the main thread, all the not yet flushed windows are pushed into the queue so that the secondary thread can process them. Finally all the output files are closed and the FEE process terminates. Once the FEE is stopped, the user can download the outputs produced with a proper button in the GUI. This triggers a shell script into the access point that prepares the outputs in an archive (using tar and gzip) and sends it to the user. It is also possible to download the output while the FEE is still active.  %In this way the data can be stored saving a consistent amount of space, thanks to the high entropy of the output produced, as shown in the results discussed in subsection \ref{subsec:results_offline}.

\begin{figure}[t]
    \begin{center}
      \includegraphics[width=\columnwidth]{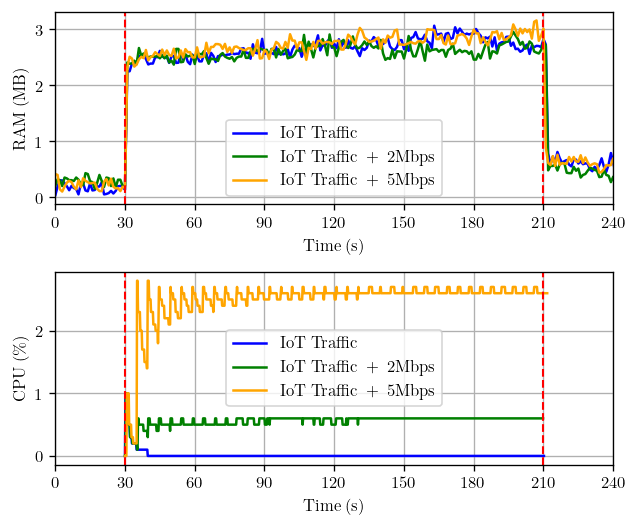}
      \caption{RAM and CPU utilization during FEE execution under different network conditions}
      \label{fig:cpu_ram_time}
    \end{center}
\end{figure}

\begin{figure}[t]
    \begin{center}
      \includegraphics[width=\columnwidth]{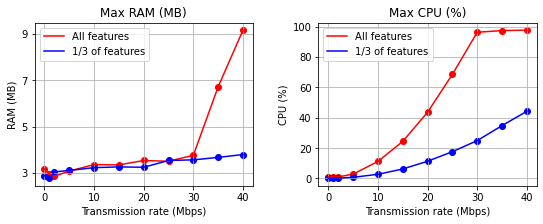}
      \caption{Maximum value of RAM and CPU used by the FEE process under different transmission rates}
      \label{fig:cpu_ram_bw}
    \end{center}
\end{figure}

\subsection{Performance evaluation}
In order to understand the impact of the proposed tool in realistic network scenarios, we installed Feature-Sniffer on a commercial Wi-Fi Access Point supporting the OpenWRT project, namely a Linksys AC3200 device. The selected Access Point is characterized by 512 MB RAM, 1.8GHz CPU and it supports the latest version of OpenWrt (21.02.0).
As a first step, we flushed the Access Point with the most up to date version of the OpenWrt operating system.
Once the operating system is up and running, we installed the Feature-Sniffer GUI as an extension of the LuCI Web interface, adding both the client (HTML/Javascript views) and server parts (Lua controller). Finally to install the Feature Extraction Engine, we compiled the C program for the selected architecture, by using the OpenWrt Software Development Kit. %with the proper target/subtarget (\texttt{mvebu/cortexa9} in our case). 
The compilation procedure produces a package to be installed into the system using the \texttt{Opkg} packet manager, a fork of \texttt{ipkg} used in OpenWrt systems.
%As the FEE is installed, the tool is ready to be used to capture network traffic features on-the-fly, as the traffic flows through the Access Point.
To evaluate the performance of Feature-Sniffer in a real-life scenario, we use the Linksys AC3200 access point to set up a Wi-Fi network in our laboratory, and let 10 IoT devices (smart bulbs, cameras, etc...) and 3 non-IoT devices associate to it. Each device is characterized by a different traffic profile: for what concerns IoT devices, we stimulate the production of traffic by interacting with them during the tests (e.g., turning on/off the smart bulbs or activating the smart cameras). Moreover, in order to test the performance of Feature-Sniffer under considerable traffic loads, we configured a Raspberry Pi3 and a laptop as \texttt{iperf3} client and server, respectively. The \texttt{iperf3} tool allows to generate different traffic profiles, controlling  the traffic bandwidth, the packet size, the transport protocol and the transmission period. 
We keep track of the CPU and RAM consumption of the FEE process (tracked by PID) using two specific shell scripts, which are able to periodically measure the values with a sampling rate of five samples per second. 
%For what concerns RAM usage, the script reads the value from the \texttt{/proc/meminfo} file, which is updated from the operating system each time memory allocation changes. As for the CPU usage, we used the Linux \texttt{ps} command to track the percentage of CPU used from the FEE process, passing the correct process identifier as argument.
Several tests have been performed, with the goal of monitoring the performance of the Feature-Sniffer tool under different traffic loads while understanding its impact on the Access Point working operations.
In particular we created one Feature-Sniffer configuration, activating all available features to be extracted for each device in the network, setting the window size to 5 seconds. To understand the difference in CPU and RAM enabling/disabling the Feature-Sniffer tool, the tests have been executed as follows: first, we start the traffic generation process with iperf3 and the RAM monitoring on the access point. After 30 seconds, we start the FEE process and the CPU monitoring process, letting them run for 180 seconds. Finally, we let other 30 seconds pass and we conclude the test.
We repeated the tests three times, each time changing the traffic rate generated by the Iperf3 client: 0 Mbps (no additional traffic), 2 Mbps and 5 Mbps. The usage of CPU and RAM for the three analyzed network cases are shown in Figure \ref{fig:cpu_ram_time}, with the red dashed lines delineating the three periods of execution of the tests. As the plots highlight, Feature-Sniffer has a minimal impact on the performance of the access point:
the RAM usage is characterized by an increase of around 2-3 MB when enabling the FEE for all the cases, independently from the iperf3 client transmission rate. For what concerns the CPU, the figure shows that different network conditions result in different CPU consumption, since the tool has more packets to analyze in each time window. In any case, the CPU usage remain limited to less than 3\%. During the simulation we have also kept track of the packet error rate as reported by the iperf3 client-server interaction: no particular differences have been observed when executing/not executing the tool.\\
To investigate the behaviour of the tool with higher traffic loads we also analyzed the maximum CPU and RAM used by the FEE process in function of the transmission rates when using two Feature-Sniffer configurations: one with all the available features and a lighter version with just 1/3 of the available features turned on (selected randomly). %The subset of features has been selected randomly among the several types of features available. 
As shown in Figure \ref{fig:cpu_ram_bw} when all features are selected, traffic loads exceeding 30Mbps cause the FEE CPU usage to saturate and in turn RAM usage to sharply increase. With the reduced set of features, both RAM and CPU usage remain at acceptable values also for very high traffic loads.

\section{Use cases}
\label{sec:application}
To showcase the benefit of using Feature-Sniffer in a realistic IoT forensic use case, we leverage it to perform analysis of IoT video cameras traffic. Such devices are particularly interesting from a forensic point of view, due to their widespread in Smart Homes / Smart Building scenarios: indeed, they constitute the perfect digital witness by-design. For privacy reasons, IoT cameras generally stream encrypted video traffic towards cloud services / mobile apps, making it impossible to retrieve the actual video content. However, recent studies have shown that some information leakages occur in encrypted IP video traffic, allowing for partial understanding of the events happening in the field of view of the camera \cite{wampler2015information,li2016side,video_classification}. In a nutshell, variations in packet size and inter-arrival times could indicate activity in a video stream, regardless of the used video codec, camera, and processing hardware.

The forensic tasks we aim to solve are (i) camera identification and (ii) activity recognition from encrypted traffic. We show that Feature-Sniffer is a formidable tool to easily collect the traffic features needed to successfully accomplish the two tasks, at the same time reducing the storage space needed for storing evidences.

\subsection{Environment setup and data collection}
%First, we installed Feature-Sniffer in a Raspberry PI 4 model B with 4GB RAM. In details, we flushed on the Raspberry the OpenWrt 21.02.0 firmware, setting it up to act as a standard Access Point connected to the Internet via Ethernet.  Then, we loaded the Feature-Sniffer functionalities: first, we installed the user interface as an application of the LuCI web interface (both client and server sides). For what concerned the Feature Extraction Engine, we first compiled the C program for the Raspberry architecture, using the corresponding OpenWrt Software Development Kit with the proper target/subtarget. Then, we installed the Feature-Sniffer package with the \texttt{Opkg} package manager, a fork of \texttt{ipkg} used in OpenWrt systems.
As a first operation we selected three Wi-Fi IoT cameras of different brands and connected them to the Access Point already introduced in subsection \ref{sec:tool}D. The cameras in consideration are: Ezviz C6N, Tapo C100, Teckin TC100.
%Finally, we connected to the access point three WiFi-enabled IoT cameras of different brands: an Ezviz C6N, a Tapo C100 and a Teckin TC100. 
All cameras have the same resolution (1920x1080) and can be activated through a smartphone application. The three cameras were mounted in an indoor space very close to each other, so that their Field of Views (FoV) were overlapping. We configured each camera to stream video over Wi-Fi continuously, disabling the audio coming from the embedded microphones. 
Then, we performed four different activities in front of the cameras, each one of the duration of 30 minutes:
\begin{itemize}[nosep, wide]
    \item \textit{No activity}: a person sitting still in the FoV of the cameras.
    \item \textit{Slow movement}: a person walking slowly and calmly in the FoV of the cameras
    \item \textit{Fast movement}: a person frenetically moving in front of the cameras
    \item \textit{Lights On/Off}: the artificial light in the indoor space is turned on or off. No other light source is present. 
\end{itemize}
In order to produce one dataset for each activity, we set up four configurations from the Feature-Sniffer GUI. Each configuration uses a 2 second window containing the following features, derived according to the study in \cite{video_classification} (58 features in total):
\begin{itemize}[nosep, wide]
    \item Downlink Packet size (8 features: all statistical values)
    \item Uplink Packet size (8 features: all statistical values)
     \item Downlink Inter-arrival time (5 features: mean, median, variance, st. deviation., kurtosis)
    \item Uplink Inter-arrival time (5 features: mean, median, variance, st. deviation., kurtosis)
    \item PMF features of Downlink packet lengths (16 features: all size ranges)
    \item PMF features of Uplink packet lengths (16 features: all size ranges)
\end{itemize}
The device filter of all configuration is set to the MAC address of the three cameras and a different ground-truth label is used for each activity-camera pair: numbers from 1 to 3 for the No Activity configuration (Ezviz=1, Tapo=2, Teckin=3), numbers from 4 to 6 for Slow Movement and so on for a total of 12 labels.
For each activity, we start the proper configuration to extract features while performing the activity in front of the cameras. After 30 minutes, we stop the configuration and download the obtained features for later analysis. The final dataset therefore contains roughly 900 observation windows per activity per camera, each one containing 58 traffic features and 1 ground truth label.

\section{Experimental Results}
\label{sec:results}
In this section we report the experimental results obtained by applying Machine Learning algorithms to the data obtained from the configurations explained in the previous section. In particular we have downloaded the features extracted from Feature-Sniffer on a laptop (Intel i7-8750 with 6 CPUs @ 2.2GHz, 16 GB of RAM) running Ubuntu 18.04.4 and Python version 3.9with the use of the scikit-learn library\footnote{https://scikit-learn.org/stable/} to apply several algorithms for several use cases.

\subsection{Camera identification}
As first task, we have trained different machine-learning classifiers (Random Forest, Adaptive Boosting, Decision Trees) with the goal of identifying the camera that is being used. For the task at hand, first we have merged the four outputs in one single CSV files, being this possible since the labels have been assigned properly to avoid conflicts. Then, the labels (1,4,7,10) in the merged dataset have been renamed as Ezviz, while the labels (2,5,8,11) and (3,6,9,12) as Tapo and Teckin, respectively. We relied on a 10-fold cross-validation process: the dataset is split in ten folds and at each iteration nine of them are used for training and one for testing. The Random Forest classifier achieved the best results with an Accuracy, Precision and Recall of 99.88\%.
\begin{figure}[h]
\centering
\begin{subfigure}[t]{0.45\columnwidth}
    \centering
    \caption{All Cameras}
    \includegraphics[width=\textwidth]{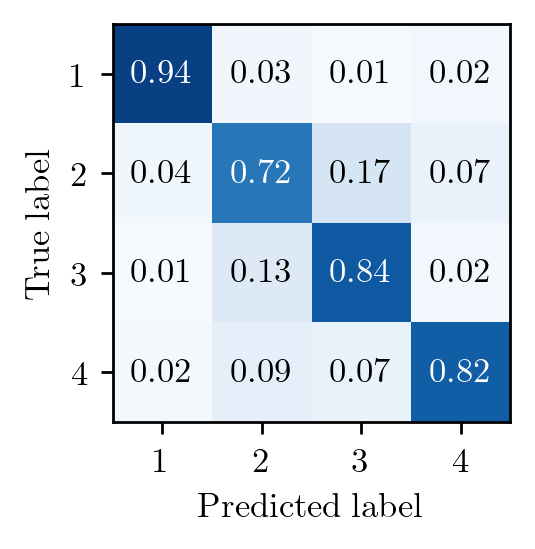}
    \label{fig:act_id}
\end{subfigure}
~ %add desired spacing between images, e. g. ~, \quad, \qquad, \hfill etc. 
  %(or a blank line to force the subfigure onto a new line)
\begin{subfigure}[t]{0.45\columnwidth}
    \centering
    \caption{Ezviz}
    \includegraphics[width=\textwidth]{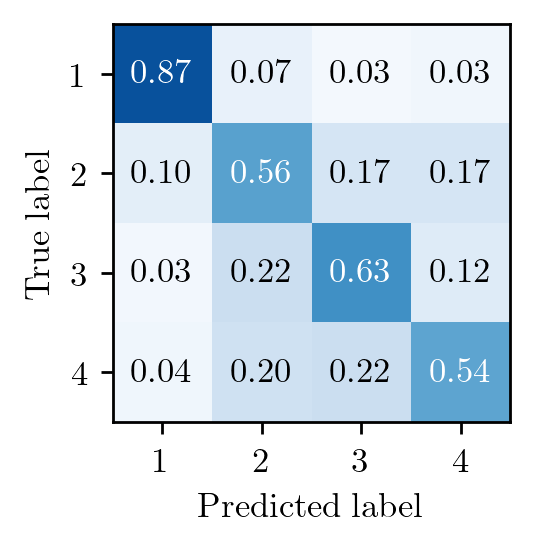}
    \label{fig:ezviz_id}
\end{subfigure}
~ %add desired spacing between images, e. g. ~, \quad, \qquad, \hfill etc. 
%(or a blank line to force the subfigure onto a new line)
\begin{subfigure}[t]{0.45\columnwidth}
    \centering
    \caption{Tapo}
    \includegraphics[width=\textwidth]{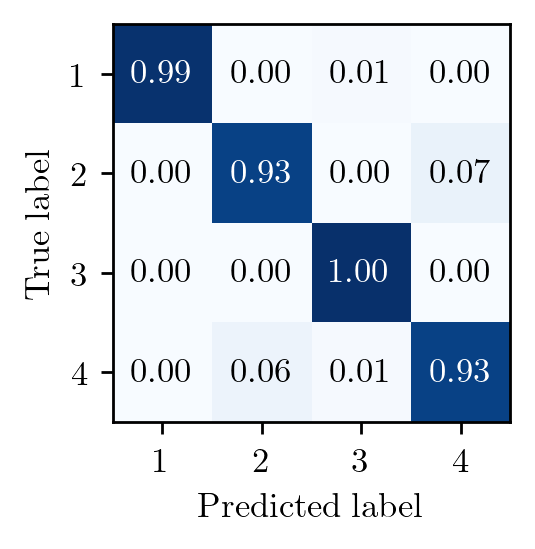}
    \label{fig:tapo_id}
\end{subfigure}
~ %add desired spacing between images, e. g. ~, \quad, \qquad, \hfill etc. 
%(or a blank line to force the subfigure onto a new line)
\begin{subfigure}[t]{0.45\columnwidth}
    \centering
    \caption{Teckin}
    \includegraphics[width=\textwidth]{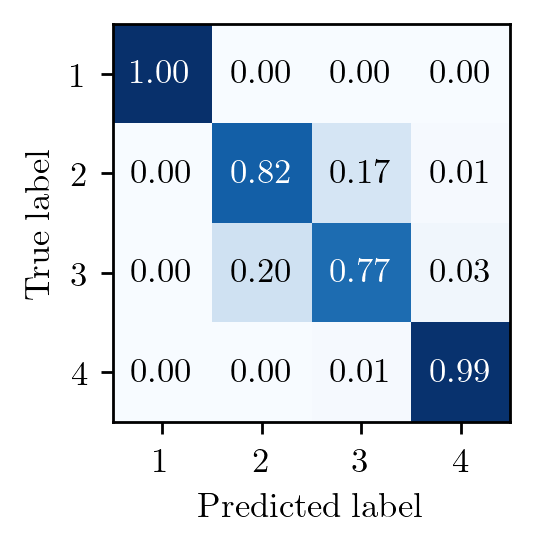}
    \label{fig:teckin_id}
\end{subfigure}
\caption{Random Forest confusion matrix for Activity Recognition using the three cameras (a) and single cameras independently (b), (c), (d)}\label{fig:all_act_id}
\end{figure}
\subsection{Activity recognition}
\label{subsec:results_live}
As a second IoT forensic task, we explore the possibility of recognizing the activity occurring in the camera FoV starting from the features extracted from the encrypted traffic. First, we assume that no classifier for camera identification is available. Therefore,
we merge the four outputs obtained from Feature-Sniffer, this time renaming the ground truth labels as it follows: (1,2,3) No Activity with label=1, (4,5,6) Slow Movement with label=2, (7,8,9) Fast Movement with label=3 and (10,11,12) Lights On/Off with label=4. Similarly to before, we rely on a 10-fold cross-validation approach to train several Machine-Learning classifiers. We report here only the best performance, obtained again with the Random Forest algorithm. Figure \ref{fig:act_id} shows the confusion matrix obtained in this case: as one can see performance are quite good, with an average F1-score of about 83\% and the majority of classification errors occurring between Slow and Fast movements. 
To improve the performance, we leverage the camera identification classifier to specialize the training procedures. In details, we assume that the recognition pipeline first identifies which camera the traffic belongs to and then tries to recognize the activity occurring in the FoV. To this end, we train three new classifiers, on the subsets of the dataset corresponding to the three different cameras, always following a 10-fold cross-validation approach. The resulting confusion matrices for the three cases are reported in Figures \ref{fig:ezviz_id}, \ref{fig:tapo_id} and \ref{fig:teckin_id}. As it can be seen from the confusion matrices, the best results have been achieved by using the Tapo camera, with average accuracy of 96.6\%, while the worst results are observed for the Ezviz camera, with an accuracy of 65\%. The Teckin camera has obtained good results with an overall activity recognition accuracy of 90\%.
\subsection{Required storage}
To highlight the benefits of using Feature-Sniffer in terms of used storage, we also captured the complete network traffic traces also using \texttt{tcpdump} to store them in standard PCAP files. The four activities of the three cameras combined in a single PCAP file result in 2,73 GB storage used, that compressed become 2.16 GB (compression ratio 1.26). Instead the Feature-Sniffer output datasets contained in a 3.02 MB CSV once compressed result in a 865 KB archive (compression ratio 3.58). In this case, Feature-Sniffer allows to perform IoT forensics analysis with $2,5\times10^3$ times less storage space than a traditional approach based on PCAP files.

\section{Conclusions}
\label{sec:concl}
This work has presented Feature-Sniffer, a tool to turn an OpenWrt based Access Point in a device able to easily collect IoT devices traffic features for forensic purposes.  To show the impact of the tool in real-life scenarios, we have installed the tool in one of the most recent and popular Access Point and performed several performance tests, highlighting the low CPU/RAM consumption of the tool. Finally, to show possible uses of the tool, we have presented two application cases with corresponding experimental results in which the output of the tool is used to train ML classifiers for different tasks.
%As future research directions, we plan to install the tool into some of the most popular commercial access points to test the performance in real scenarios, as well as considering more features and more use cases to build working ready-to-use
As future research directions, we plan to focus our research in further reducing the space required for feature storage using lossy compression techniques, also analyzing the trade-off existing between the space needed to store the features and the accuracy of the subsequent traffic analysis tasks.
%In addition, we plan to focus our research on feature storage, studying both lossless and lossy network feature compression algorithms.
%$Finally we plan also to include the inference procedure at the access point level in order to detect real-life events as soon as they occur.

%Finally, we plan to expand the work currently done by enabling Feature-Sniffer with inference capabilities, bringing the inference procedure to the Access Point level and hence detecting real-life events from the traffic features as soon as they occur.

\addcontentsline{toc}{chapter}{Bibliografy}
\bibliographystyle{IEEEtran}
\bibliography{bibl_tesi}

\end{document}